\documentclass[12pt]{article} 
\usepackage{english}
\newcommand{\bee}{\begin{equation}}                                             
\newcommand{\ee}{\end{equation}}                                                
\newcommand{\ba}{\begin{array}}                                                 
\newcommand{\ea}{\end{array}}                                                   
\newcommand{\bea}{\begin{eqnarray}}                                             
\newcommand{\eea}{\end{eqnarray}}                                               
\begin{document}                                                                
\thispagestyle{empty}                                                           
\begin{flushright}                                                              
MPI-PhT 96-78  \\
August 1996                                                        
\end{flushright}                                                                
\bigskip\bigskip                                                                
{\it Comment on}                                                                
\begin{center}                                                                  
{\bf \Large{The two-phase issue in the $O(n)$ non-linear 
}}\vskip2mm                
{\bf \Large {$\sigma-$model: a Monte-Carlo study}}
\vskip 2mm by B.All\'es, A.Buonanno and G.Cella               
\end{center}                                                                    
\vskip 1.0truecm                                                                
\centerline{\bf                                                                 
Adrian Patrascioiu}                                                             
\vskip5mm                                                                       
\centerline{Physics Department,                                                 
University of Arizona, Tucson, AZ 85721, U.S.A.}                                
\vskip5mm                                                                       
\centerline{and}                                                                
\vskip5mm                                                                       
\centerline{\bf Erhard Seiler}                                                  
\vskip5mm                                                                       
\centerline{Max-Planck-Institut f\"ur                                           
 Physik}                                                                        
\centerline{ -- Werner-Heisenberg-Institut -- }                                 
\centerline{F\"ohringer Ring 6, 80805 Munich, Germany}                          
\vskip 2cm                                                                      
                                                                                
In their recent paper \cite{alles} Alles et al present new numerical
evidence favoring the absence of a massless phase in the $O(N)$ nonlinear
$\Sigma$ models with $N>2$. While we have nothing to say about their
numerics,
we would like to repeat our objections to some of the statements made by
Alles et al:

1) `The results for $O(8)$ support the asymptotic freedom scenario.'

As we have stressed repeatedly in the past \cite{1/N}, Kupianen
\cite{Kupi}  proved rigorously
that the $1/N$ expansion produces the correct asymptotic expansion at
fixed
$\beta~=\beta/N$ and the only issue is whether the expansion is uniform in
$\beta~$. Since it is known rigorously that the spherical model has
$\eta=0$, if a nonuniformity in $\beta~$ does in fact exist, to see any
deviations from $\eta=0$ one would have to probe larger and larger values
of $\beta~$ as one increases $N$. Therefore if one wishes to determine the
universality class of the $O(N)$ models with $N>2$ (as revealed by the
value
of $\eta$), one should investigate $O(3)$ not $O(8)$, where the true
asymptotic value may emerge only at huge correlation length.

2) `Assuming finite-size scaling (FSS), it has been shown that $O(3)$
presents
asymptotic scaling starting from $\xi=10^5$.'

As we stated in our Comments to the Kim and Caracciolo et al papers
\cite{PS}, contrary to their claims, these authors have not established
the existence of asymptotic scaling in $O(3)$, but in fact implicitly
preassumed it. Indeed,
it is again a rigorous fact that perturbation theory (PT) in $1/\beta$  
at fixed lattice size $L$ gives the correct asymptotic expansion and the  
only open issue is whether this expansion is uniform in $L$. In their FSS
investigations, Kim and Caracciolo et al assumed that there exists a    
$\beta$ independent $L_{min}$ on which one can apply FSS. The existence of
such an $L_{min}$ is equivalent to the assumption that the model is      
asymptotically free, since this is a true property of PT and the latter
is surely valid at fixed $L=L_{min}$. As we showed in a recent paper
\cite{PLB}, the corrections to FSS, although quite small are there and
thus the claims of Kim and Caracciolo et al are questionable.

3) `The $O(3)$ model with Symanzik action does not show KT behavior.'

The Symanzik action was invented precisely to improve the agreement of 
lattice PT with continuum PT and the latter does produce $\eta=0$. The 
price is
the introduction of an anti-ferromagnetic coupling, whose effect could  
be a lack of monotonicity of certain thermodynamic variables.
Consequently it is hard to know what the small variation observed by   
Alles et al in $R_{KT}$ is supposed to indicate. On the other hand, the
constancy of $R_{KT}$ with precisely $\eta=1/4$ and the clear drop of
$R_{PT}$
found by us for the model with standard action seems at the very least
intriguing, if in fact the asymptotic freedom scenario is the correct
one. To allow the reader to form a better impression whether for the
Symanzik action the KT or the PT scenario looks more likely,
Alles et al should display (on the same scale) both $R_{KT}$ and
$R_{PT}$.

\end{document}